\documentclass[%
aip,
amsmath,amssymb,
reprint,%
]{revtex4-2}

\usepackage{color}
\usepackage{graphicx}% Include figure files
\usepackage{dcolumn}% Align table columns on decimal point
\usepackage{bm}% bold math
\begin{document}

\preprint{APS/123-QED}

\title{Persistence of hydrodynamic envelope solitons:\\
	detection and rogue wave occurrence}% Force line breaks with \\

\author{Alexey Slunyaev}
\thanks{corresponding author}
\email{Slunyaev@appl.sci-nnov.ru}
\affiliation{%
 HSE University, N. Novgorod, Russia\\
 Institute of Applied Physics of the Russian Academy of Sciences,  N. Novgorod, Russia\\
 Nizhny Novgorod State Technical University n.a. R.E. Alekseev,  N. Novgorod, Russia
}%

\date{\today}% It is always \today, today,

\begin{abstract}
The observation of a wave group persisting for more than 200 periods in the direct numerical simulation of nonlinear unidirectional irregular water waves in deep water is discussed. The simulation conditions are characterized by parameters realistic for broad-banded waves in the sea. Through solution of the associated scattering problem for the nonlinear Schr\"odinger equation, the group is identified as the intense envelope soliton with remarkably stable parameters. 
Most of extreme waves occur on top of this group, resulting in higher and longer rogue wave events.  
%\begin{description}
%\end{description}
\end{abstract}

\maketitle

%\tableofcontents

\section{\label{sec:Introduction}Introduction}

The intriguing problem of anomalously high sea waves (rogue waves) is in the focus of numerous multifaceted studies, see e.g. reviews \cite{Dystheetal2008,Kharifetal2009,Onoratoetal2013}. The approach to study this complicated natural phenomenon by means of the direct numerical simulation (DNS) of the irregular wave evolution in time has become popular, thanks to the high performance of modern computers and improved numerical algorithms. The potential Euler equations serve well when simulating the wind-generated gravity waves on the sea surface. They may be solved by existing numerical algorithms incomparably quicker than the Navier -- Stokes equations and the spectral Zakharov equations \cite{Tanaka2001,Chalikov2016}, what allows the simulation of large wave ensembles within strongly- or fully nonlinear frameworks. A series of the research works based on the DNS of the Euler equations in 2D and 3D geometries has been completed in the recent years, see e.g. \cite{Ducrozeetal2007,Xiaoetal2013,BitnerToffoli2014,Brennanetal2018} among many others. 

In our previous research  \cite{SergeevaSlunyaev2013,Slunyaevetal2016,SlunyaevKokorina2017} we have performed the stochastic numerical simulation of waves on the water surface with spectral parameters typical for oceanic wind waves using the High Order Spectral Method \cite{Westetal1987}. These simulations were performed  with the purpose to better understand the mechanisms of generation of abnormally large waves, and to evaluate rogue wave characteristics, including the probability distribution function (PDF). In the simulations of unidirectional irregular deep-water waves extremely long-living wave groups were observed, which were related in Ref. \cite{SlunyaevKokorina2017} to the strongly nonlinear counterpart of the envelope solitons of the nonlinear Schr\"odiner equation. No firm justification of this conjecture was provided though.

Very short long-living groups in infinitely deep water with the maximum steepness close to the wave breaking onset were first observed in the numerical simulations of the primitive hydrodynamic equations in \cite{DyachenkoZakharov2008}, and later in \cite{Slunyaev2009}. Such strongly nonlinear hydrodynamic envelope solitons were later on reproduced in laboratory conditions: single solitons first \cite{Slunyaevetal2013PhysFluids} and then interacting pairs of solitons \cite{Slunyaevetal2017}. 
However, these soliton groups were propagating on the surface of calm water and were not surrounded by other waves.
Meanwhile the concept of soliton gas in hydrodynamics is discussed in recent publications with different level of rigor \cite{Osborneetal2019,Redoretal2019,Suretetal2020}. 

In the present work we reconsider our observation of long-living wave groups in the numerical simulation of irregular sea waves and discuss their nature with solid grounds. We apply the nonlinear analysis based on the Inverse Scattering Transform (IST) \cite{Slunyaev2006,Slunyaev2018} which helps to reveal soliton-type nonlinear groups and to evaluate them quantitatively.

The IST was probably first used for the analysis of the soliton content of irregular water waves in the works \cite{Osborneetal1991,OsbornePetti1994} dealing with experimental data under the condition of shallow water. The potential of application of the IST to oceanic waves was discussed in detail in the book by A.~Osborne \cite{Osborne2010}, and in the series of publications \cite{Osborneetal2005,IslasSchober2005,Schober2006,BruhlOumeraci2016,Suretetal2020}; it was also applied to optical pulses, see e.g. \cite{Randouxetal2016,MullyadzhanovGelash2019}. The IST formulated in periodic interval was used in all the mentioned above studies except \cite{MullyadzhanovGelash2019, Suretetal2020}. In all these works the location of a particular soliton can be in principle found by analyzing the eigenfunctions of the associated scattering problem, what is a demanding task. Following our original approach \cite{Slunyaev2006}, the application of a sliding window transformation allows to better adjust the local carrier, to locate the solitons and to consider the scattering problem in infinite line for compact potentials. As we show below, these features provide with surprisingly robust results of the IST-based analysis when applied to evolving strongly nonlinear strongly modulated wave trains. 
%The interpretation of the results is straightforward.

The paper is organized as follows. The conditions of the numerical simulation when the long-living wave groups were observed are discussed in Sec.~\ref{sec:StochasticSimulations}. The principle scheme of the IST-based nonlinear analysis is explained in Sec.~\ref{sec:IST}; it is applied to a toy example of a single hydrodynamic envelope soliton. The main part of the work is presented in Sec.~\ref{sec:Results}, where the numerically simulated evolution of strongly nonlinear unidirectional water waves is interpreted with the help of the windowed IST analysis. We conclude the paper with a discussion of the significance of the results and of the perspectives in Sec.~\ref{sec:Conclusions}.

\section{\label{sec:StochasticSimulations}Direct numerical simulation of irregular unidirectional waves in deep water}

The simulation of planar surface waves with one horizontal axis $Ox$ is discussed in this paper. The initial condition is specified according to the JONSWAP-shape power spectrum \textcolor{black}{\cite{Hasselmannetal1973}}
\begin{align} \label{JONSWAPShape}
	S(\omega) \sim \left( \frac{\omega}{\omega_p} \right)^{-5} \exp{\left[-\frac{5}{4} \left(\frac{\omega}{\omega_p}\right)^{-4}\right]} \gamma^r, \\
	r={\exp{\left[-\frac{1}{2\delta^2}\left(\frac{\omega-\omega_p}{\omega_p}\right)^2\right]}} \nonumber, \\
	\delta=
	\begin{cases}
		0.07 \quad \text{if} \quad \omega<\omega_p \\
		0.09 \quad \text{if} \quad \omega>\omega_p,
	\end{cases} \nonumber 
\end{align}
with the prescribed peak wave period, $T_p=2\pi/ \omega_p = 10$~s, significant wave height, $H_{1/3} \approx 4 \sigma \approx 3.5$~m (where $\sigma$ is the standard deviation of the surface displacement), and the spectrum peakedness parameter, $\gamma = 3$. The frequency spectrum (\ref{JONSWAPShape}) is transformed to the wavenumber spectrum $S(k)$ using the dispersion law for infinitely deep water,
\begin{eqnarray}
	\omega^2 = gk , \label{DispersionRelation}
\end{eqnarray}
where $g$ is the acceleration due to gravity.

The wave evolution in time is calculated by means of the High Order Spectral Method (HOSM, \cite{Westetal1987}). Basically, the evolution of the wave sequence is solved for $120T_p$ in a $10$-km domain, what corresponds to $60$ dominant wave lengths $\lambda_p = 2\pi/k_p$, where $k_p$ is related to $\omega_p$ according to (\ref{DispersionRelation}). Periodic boundary conditions apply. The statistical database is represented by many, $N_R > 100$, repetitions of the numerical experiment with the same initial power spectrum $S(k)$, each corresponds to a realization of random wave phases of the initial condition. In our numerical experiments we store the simulated wave data with high resolution in both, space and time, and hence are able to examine the spatio-temporal wave evolution in detail. 

The time series at a few, $N_L \ge 1$, equally spaced locations within the simulated spatial domain are retrieved and used for the further statistical analysis. Accordingly, the total number of the processed time series is $N_R N_L$. The number $N_L$ may take the value from $1$ up to the number of grid points along the $Ox$ axis, which was as much as $N_x = 2048$. One should understand that in the case of too dense locations of the ``measurement'' points (i.e., when $N_L$ is too large), the time series become correlated what can lead to spurious artifacts in the statistics.
 
\begin{figure}[b]
	\includegraphics[width=8cm]{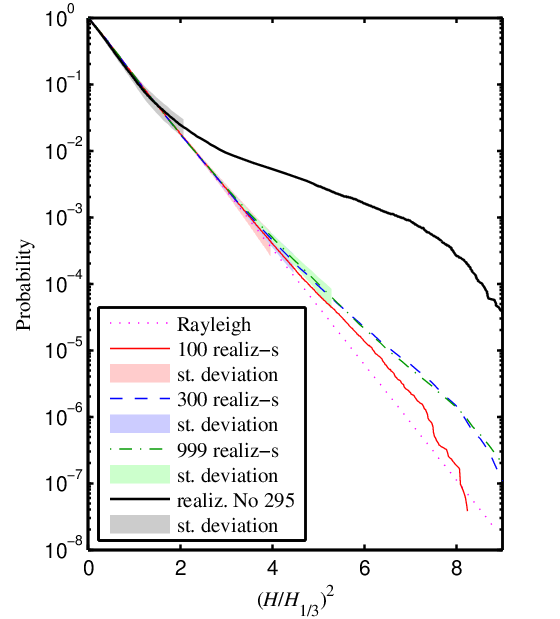}% Here is how to import EPS art
	\caption{\label{fig:PDF} Exceedance probability distribution functions calculated in the direct numerical simulation for different number of realizations (see $N_R$ in the legend). The PDF for a particular realization No 295 is plotted by the thick solid curve. The Rayleigh function is given by the dotted straight line (note that the horizontal axis corresponds to the squared scaled wave height).}
\end{figure}

When analyzing the collected data, we noticed in \cite{SlunyaevKokorina2017} that though the wave height probability distribution function averaged over the large number of realizations agreed rather well with the celebrated Rayleigh distribution (see Fig.~\ref{fig:PDF}), the PDFs plotted for some particular realizations demonstrated extraordinary behavior (see realization No 295 in Fig.~\ref{fig:PDF}). 
In the figure $H$ denotes the wave height, calculated from the time series according to the zero-crossing method. 
The significant wave height $H_{1/3}$ is defined as the mean among the one-third of the highest waves. 

The thin curves in Fig.~\ref{fig:PDF} represent the wave height exceedance probability distributions for $N_R = 100$, $N_R = 300$ and $N_R = 999$ realizations (see the legend), when the time series from all the grid points are used (i.e., $N_L=N_x$). Note that the curves for $N_R = 300$ and $N_R = 999$ practically coincide within the range $H/H_{1/3} < 3$ and clearly exhibit increase of the probability above the Rayleigh reference curve starting from the wave height excess of about $H/H_{1/3} = 2$.
The filled areas show the estimate of the confidence interval calculated as the standard deviation among the PDFs obtained for the time series retrieved at different $N_x$ locations (i.e., $N_L=1$ for the every subset), see details in \cite{SlunyaevKokorina2017}. 
The PDF for the particular realization No 295 is shown with the thick solid curve; it is well above the averaged distribution functions, demonstrating much higher probability of large waves.

\begin{figure}[b]
	\includegraphics[width=9cm]{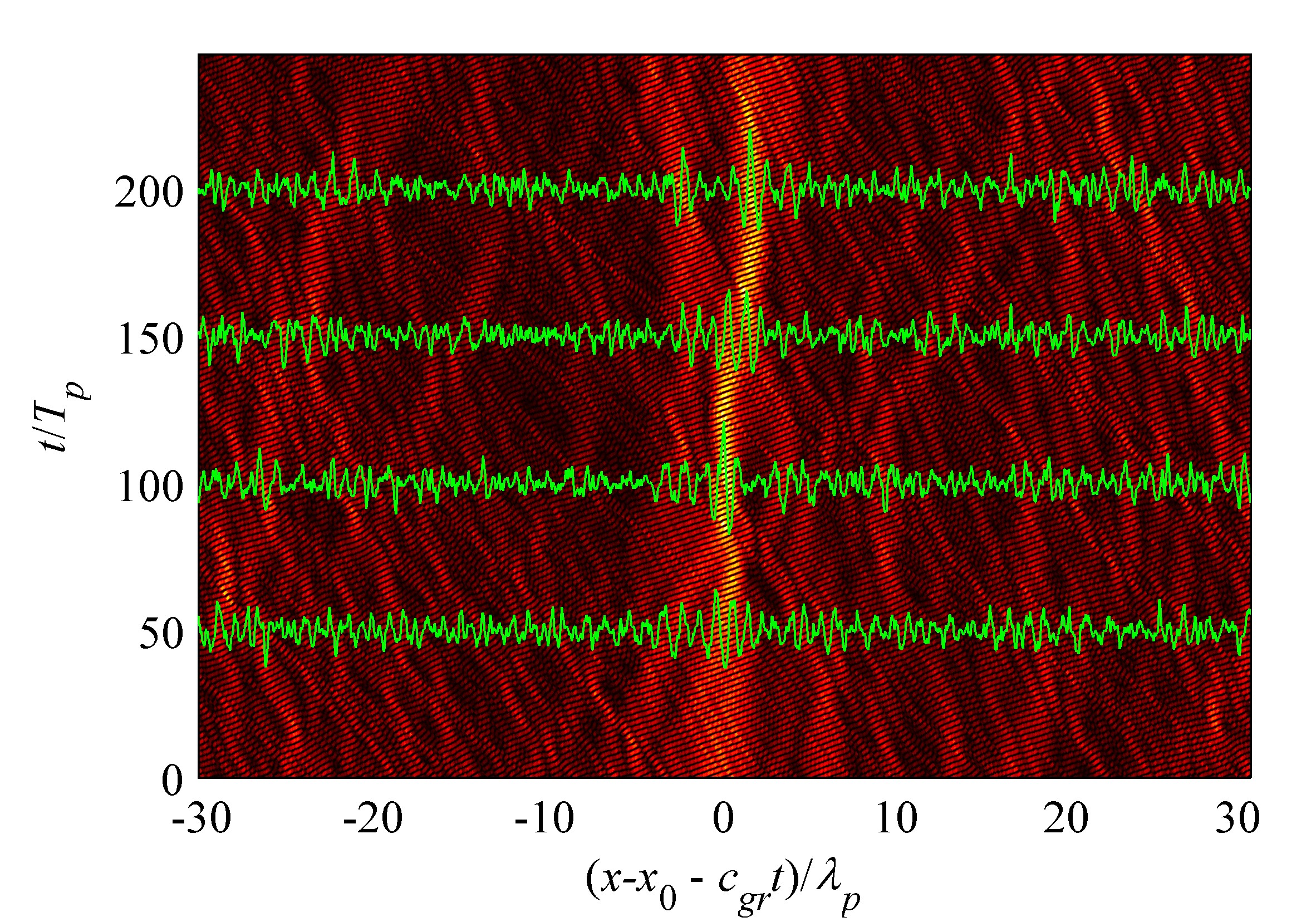}% Here is how to import EPS art
	\caption{\label{fig:Surface1} False color representation of the evolution of the sea surface in the experiment No 295, and a few snapshots of the surface (green curves). The comoving frame of reference is used.}
\end{figure}

The thorough investigation of the realization No 295 reveals the mechanism of occurrence of the abnormal PDF. In this and a few other realizations intense wave groups are formed at the early stage of the evolution. They preserve their group structure for surprisingly long. The sea surface simulated in the series No 295 for a longer period of $240T_p$ is shown in Fig.~\ref{fig:Surface1}. A few samples of the momentary surface displacements are also plotted by the green curves. The intense group (close to the coordinate origin) may be easily discerned for more than 200 dominant wave periods $T_p$. The reference moving with the group velocity of linear waves, $c_{gr} = \omega_p/k_p/2$, is used in the figure. 

It was speculated in \cite{SlunyaevKokorina2017} that such a group should be a strongly nonlinear analogue of the envelope soliton known within the framework of the nonlinear Schr\"odinger (NLS) equation.  It is well known that within the NLS theory solitons can emerge from weakly modulated trains at the late stage of development of the modulational (Benjamin--Feir) instability. 

At the same time, waves in the the discussed numerical experiment are characterized by broad JONSWAP spectrum and moderate steepness, $k_p H_{1/3}/2 \approx 2k_p \sigma \approx 0.07$.
These conditions should mainly correspond to modulationally stable waves.
%These conditions correspond to the situation well below the threshold of modulationally unstable waves. 
The Benjamin -- Feir index $BFI$ \cite{Onoratoetal2001,Janssen2003} 
\begin{eqnarray} 
BFI = \frac{\sqrt{2} k_p\sigma}{\delta_\omega}, \label{BFI}\\
\delta_\omega = \frac{1}{\sqrt{\pi} Q_p}, \quad
Q_p=\frac{2}{\sigma^4} \int{\omega S^2(\omega) d\omega}, \nonumber
\end{eqnarray}
where the spectral bandwidth $\delta_\omega \approx 0.19$ is calculated using the Goda parameter $Q_p$ as suggested in \cite{Serioetal2005}, gives the value $BFI \approx 0.27$. 
Recall that modulationally unstable waves require $BFI>1$.
Therefore the growth of a wave modulation, and the emergence of a soliton-like group are not ordinary events. 
Up to $7$-wave interactions are taken into account in the simulation (the HOSM nonlinearity parameter was $M = 6$). Hence the solution is almost fully nonlinear \cite{Clamondetal2006}, and the persistence of a nonlinear intense group surrounded by irregular waves for more than 200 wave periods is another extraordinary event. 

During this time, the group passed the simulation domain twice. The persisting group obviously leads to statistically correlated time series of the surface displacement $\eta(t)$ collected at any different locations in the domain, even if they are distant. As follows from the snapshots shown in Fig.~\ref{fig:Surface1}, the intense group consists of just a few wave cycles. Is it indeed a soliton?

In Sec.~\ref{sec:Results} we analyze the realization No 295 by means of the own method of evaluation of the soliton content suggested in \cite{Slunyaevetal2005,Slunyaev2006} and further improved in \cite{Slunyaev2018}. The method employs the Inverse Scattering Transform for the NLS equation, formulated on infinite line and allows estimation of the amplitudes, velocities and locations of the soliton groups. The method is briefly described in the next section.

\section{\label{sec:IST}Windowed IST analysis}

The windowed IST analysis (hereafter, WIST) which may be employed to study time series or spatial series was outlined in \cite{Slunyaev2006}. Its further development was presented in \cite{Slunyaev2018}. In the present paper we apply it to the spatial series (snapshots) of unidirectional waves in deep water. To this end we assume that the focusing nonlinear Schr\"odinger equation 
\begin{eqnarray} \label{NLS}
	i \left(\frac{\partial A}{\partial t} + c_{gr} \frac{\partial A}{\partial x} \right) + \frac{\omega_0}{8k_0^2} \frac{\partial^2 A}{\partial x^2}+ \frac{\omega_0 k_0^2}{2} |A|^2 A = 0
\end{eqnarray}
can serve as the local first approximation to real water waves. Here $k_0>0$ stands for the carrier wavenumber and $\omega_0$ is the cyclic frequency; the latter are linked according to (\ref{DispersionRelation}); $c_{gr} = \omega_0/k_0/2$ is the linear group velocity of the carrier. The complex amplitude $A(x,t)$ is related to the surface displacement $\eta(x,t)$ according to the formulas which take into account three asymptotic orders,
\begin{eqnarray} \label{ReconstructionFormulas}
\eta(x,t) = \overline{\eta}+\eta^{(1)}+\eta^{(2)}+\eta^{(3)}, \\
\eta^{(1)}=\operatorname{Re}{(AE)}, \quad
\eta^{(2)}=\frac{k_0}{2} \operatorname{Re}{\left(A^2E^2\right)}, \nonumber \\
\eta^{(3)}=-\frac{1}{2} \operatorname{Im}{\left(A\frac{\partial A}{\partial x}E^2\right)}+\frac{3k_0^2}{8} \operatorname{Re}{(AE)}, \nonumber \\ \overline{\eta}=\frac{1}{4}\hat{\cal{H}}{\frac{\partial |A|^2}{\partial x}}, \quad E = \exp{(i\omega_0t-ik_0x)}. \nonumber
\end{eqnarray}
Here $\hat{\cal{H}}$ denotes the Hilbert transform, see details in \cite{Trulsen2006,Slunyaevetal2014}. The NLS theory implies the assumption of a small wave steepness, $k_0|\eta| \ll 1$, and of a narrow spectrum, $\delta_k \ll 1,$ where $\delta_k$ is the dimensionless width of the wavenumber spectrum,  $\delta_k = 2 \delta_\omega$ due to the dispersion law (\ref{DispersionRelation}).
%; the following relation holds, $\delta_k = 2\delta_\omega$, thanks to the dispersion relation (\ref{DispersionRelation}) .

The equation (\ref{NLS}) may be further reduced to the dimensionless form
\begin{eqnarray} \label{DimensionlessNLS}
	i \frac{\partial Q}{\partial T} + \frac{\partial^2 Q}{\partial X^2}+ 2|Q|^2 Q = 0
\end{eqnarray}
after the change of variables
\begin{eqnarray} \label{Nondimensionalizing}
X=\sqrt{2} s_0k_0 \left(x-c_{gr}t\right), \quad
T=s_0^2\frac{\omega_0}{4}t, \quad
Q=\frac{k_0}{s_0}A,
\end{eqnarray}
where we choose the reference steepness $s_0$ to be specified by the maximum amplitude of the complex envelope, $s_0 = k_0 \max{|A|}$. Then the modulus of the new function $Q(X,T)$ is limited by $1$.

The celebrated envelope soliton solutions for the equation (\ref{DimensionlessNLS}), which are eternally stable localized wave groups, read
\begin{eqnarray} \label{Soliton}
	Q(X,T)= \nonumber \\
	=a_s \frac{\exp{\left( i\phi_s + i \left( \left[ a_s^2-\left(\frac{v_s}{2}\right)^2\right] T + \frac{v_s}{2} X \right) \right)}}{\cosh{\left[ a_s \left( X-v_sT-X_s \right)\right]}}.
\end{eqnarray}
Here $a_s$ and $v_s$ are the dimensionless amplitude and velocity of the soliton, $X_s$ and $\phi_s$ are the location and complex phase of the soliton respectively. 

If envelope solitons are present in the given wave field $Q(X)$, then their amplitudes and velocities may be recovered from the solution of the eigenvalue problem on the complex-valued eigenvalues $\mu$ and functions $\psi_1(X)$ and $\psi_2(X)$ \cite{Ablowitzetal1974}, 
\begin{eqnarray} \label{AKNS}
\frac{d}{dX} 
\begin{pmatrix}
	\psi_1\\ \psi_2
\end{pmatrix}
=
\begin{pmatrix}
	\mu & Q(X)\\
	-Q^{*}(X) & -\mu
\end{pmatrix}
\begin{pmatrix}
	\psi_1\\ \psi_2
\end{pmatrix},
\end{eqnarray}
\begin{eqnarray} \label{mu}
\mu=\frac{a_s}{2}+i\frac{v_s}{4}.
\end{eqnarray}
The discrete eigenvalues $\mu$, which correspond to the eigenfunctions which decay when $|X| \rightarrow \infty$, specify amplitudes $a_s$ and velocities $v_s$ of the envelope solitons, though their locations and phases are determined by the structure of the eigenfunctions. 

Restricting our attention to finite potentials $Q(X)$ with compact support and to the discrete spectrum, the problem on infinite line, $- \infty < X < \infty$, may be approximated by the one in a finite interval, $X_L \le X \le X_R$. The corresponding boundary conditions are formulated as follows. The condition on the left edge of the interval is set $\psi_1(X_L)=1$, $\psi_2(X_L)=0$; the eigenvalues $\mu$ (one or several) should provide with $\psi_1(X_R)=0$ \textcolor{black}{and $\psi_2(X_R)=1$} at the right edge of the interval. As a result, the boundary-value problem may be efficiently solved numerically (we employ the shooting method). 
The dimensional amplitudes, $A_s$, and velocities, $V_s$, of the envelope solitons are obtained by inverting the transformations (\ref{Nondimensionalizing}),
\begin{eqnarray} \label{SolitonParameters}
A_s = \frac{2s_0}{k_0} \operatorname{Re}{(\mu)}, \quad
V_s = c_{gr}+\frac{s_0\omega_0}{\sqrt{2}k_0} \operatorname{Im}{(\mu)}.
\end{eqnarray}
\textcolor{black}{Note that the NLS theory does not contain any nonlinear correction to the group velocity. If a soliton is characterized by the same wave length as the carrier wave (i.e., $v_s=0$ in (\ref{Soliton})), then $\operatorname{Im}{(\mu)}=0$ and $V_s=c_{gr}$.}

As was mentioned above, the NLS theory implies the narrowband condition, what is hardly fulfilled in the real sea. At the same time waves in deep water are known to form groups, therefore it may be expected that within some short spatial intervals the local wavelengths of energetic waves are relatively similar. Then the NLS model should be more efficient in approximating the wave surface within small areas. 

It may be realized, that the estimation of the soliton content  strongly depends on the choice of the carrier wavenumber $k_0$. The similarity parameter of the NLS equation (\ref{NLS}) (i.e., the ratio of the nonlinear term over the term of dispersion)  is in fact proportional to $BFI^2$, which for the given wavenumber spectrum width $\Delta k$ and standard deviation $\sigma$ reads $BFI = 2\sqrt{2} k_0^2 \sigma/ \Delta k$, where $\delta_k = \Delta k / k_0$. Hence, $BFI$ is proportional to the squared carrier wavenumber, and the similarity parameter is proportional to $k_0^4$. 

The carrier wavenumber is not well-defined in the situation of a broad wave spectrum. 
On the other hand, the accuracy of estimation of the carrier wavenumber $k_0$ in shorter samples decreases. This problem was in the focus of our paper \cite{Slunyaev2018}, where different ways to estimate the parameter $k_0$ were examined. 
In the present work we choose $k_0$ to be equal to the mean wavenumber calculated for the power Fourier transform of the given short sample of the spatial series. The intrinsic wavenumber of soliton may differ from $k_0$ due to the imaginary part of the corresponding eigenvalue $\mu$, see (\ref{mu}) and (\ref{SolitonParameters}). 

The use of a sliding window along the $x$ coordinate (boxcar transform) allows us to locate the detected soliton groups as described below. 
From the practical point of view, only solitary groups which possess relatively large amplitudes are of the interest. Envelope solitons with larger amplitudes are narrower (see (\ref{Soliton})), therefore the minimum width of the intense soliton may be calculated. This allows us to estimate from above the necessary width of the sliding window, $l_w=x_R-x_L$. Here $x_L$ and $x_R$ are the dimensional coordinates of the left and right edges of the sample. In the analysis we define $l_w$ according to the formula
\begin{eqnarray} \label{Lw}
l_w=\frac{W}{\sqrt{2} s_0k_0},
\end{eqnarray}
where $W$ is a number of the order of 20 (see details in \cite{Slunyaev2018}), and $s_0 = k_0 \max{|A|}$ is the maximum wave steepness in the wave sample. 

\begin{figure}[b]
	\includegraphics[width=8.5cm]{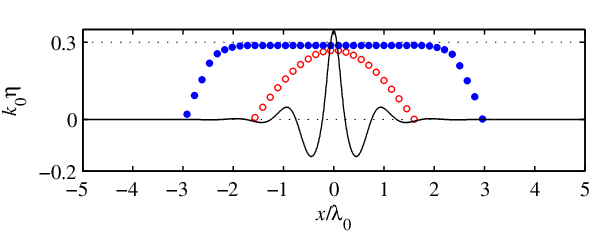}% Here is how to import EPS art
	\caption{\label{fig:EnvelopeSoliton} Envelope soliton with the steepness $k_0 \max{|A|} = 0.3$ (the solid line) and the soliton amplitudes $A_s$ detected using the windowed IST-based procedure with (empty red circles) and without (full blue circles) the \textcolor{black}{Hann} mask, $W = 20$.}
\end{figure}

In Fig.~\ref{fig:EnvelopeSoliton} the windowed IST procedure is applied to a snapshot of an envelope soliton. The surface displacement $\eta(x)$ (shown with the solid line is calculated from $Q(X)$ (\ref{Soliton}) according to (\ref{Nondimensionalizing}) and (\ref{ReconstructionFormulas}). The maximum steepness of the group may be roughly estimated based on the complex envelope $A(x)$ as $s_0 = 0.3$, though the amplitude of the wave crest $(k_0 \max{\eta})$ is in fact larger due to the nonlinear harmonics introduced by the reconstruction formulas (\ref{ReconstructionFormulas}).

The windowed transform splits the long spatial series into overlapping shorter samples of the length $l_w$. A sample $\eta_w(x)$, $x_L \le x \le x_R$, is the only input data which is used by the procedure of the IST analysis. At first, the complex amplitude $A_w(x)$ is calculated with the help of an iterative procedure which inverts the relation (\ref{ReconstructionFormulas}), similar to \cite{Trulsen2001,Slunyaevetal2014}, using the local mean wavenumber as the value of $k_0$. Then $A_w(x)$ is transformed to $Q_w(X)$ using (\ref{Nondimensionalizing}), and after that the eigenvalue problem (\ref{AKNS}) with appropriate boundary conditions is solved to obtain the discrete eigenvalues $\{\mu \}$. The soliton parameters $\{ A_s \}$ and $\{ V_s \}$ are calculated for the given sample as a result. 

The values of $A_s$ obtained in the sliding window are shown with symbols in Fig.~\ref{fig:EnvelopeSoliton} as the function of location of the window, $x_w = (x_L+x_R)/2$. The estimated soliton amplitude reaches the maximum when the window covers the entire soliton group. The soliton amplitudes decay when lesser part of the soliton is captured by the window. Consequently, the soliton amplitude as the function of coordinate, $A_s(x_w)$, forms an arc, see Fig.~\ref{fig:EnvelopeSoliton}. When the sliding window is significantly longer than the soliton width, the estimated soliton amplitudes in the neighboring samples have similar values and form a plateau of the arc (the full blue circles in Fig.~\ref{fig:EnvelopeSoliton}). The value of $A_s$ achieved in the central part of the ‘arc’ is used as the estimate of the true amplitude of the envelope soliton. The location of the soliton may be better determined using a slightly smaller value of $W$ and/or by applying the Hanning mask which makes the procedure more stable (the empty red circles in Fig.~\ref{fig:EnvelopeSoliton}). 

The actual procedure of the WIST which is applied in this work consists of two steps. A shorter window with the  \textcolor{black}{Hann smoothing mask $M(x)=0.5 (1+\cos{\left( 2\pi (x-x_w)/l_w \right)})$} is used at the first stage to locate envelope solitons. At the second stage the problem is solved more precisely for the selected solitons among the largest revealed soliton groups. At the second stage a wider window of a rectangular shape is used.
The windowed IST procedure was tested in \cite{Slunyaev2018} on the example of soliton-type solutions of the primitive Euler equations. It was shown that the soliton amplitudes $A_s$ obtained using this procedure are equal to the actual envelope amplitudes, $A_{env} = (A_{cr} + A_{tr})/2$, with the accuracy of about 10-15\% for the range of steepness $k_pA_{env} < 0.3$. Here $A_{cr}$ and $A_{tr}$ are the maximum amplitudes of the crest and of the trough respectively of the waves which belong to the soliton group. The soliton group shown in Fig.~\ref{fig:EnvelopeSoliton} consists of about two wave cycles; its steepness $k_pA_{env}$ is about $0.3$. Therefore the result of the tests reported in \cite{Slunyaev2018} seems to be unexpectedly good bearing in mind the implied assumptions and approximations. 
The advantage of the IST-based analysis is that it can disclosure envelope solitons even when they are completely disguised by surrounding random waves. 

In the next section the described nonlinear analysis is applied to the snapshots of evolving irregular waves shown in Fig.~\ref{fig:Surface1}. The WIST procedure is fully automated; it uses the spatial series of the surface displacement as the only input data.

\section{\label{sec:Results}Revealed hydrodynamic envelope solitons}

The water surface evolution which is analyzed using the WIST, is shown in Fig.~\ref{fig:Surface1}; details of the numerical simulation are described in Sec.~\ref{sec:StochasticSimulations}, see also \cite{SlunyaevKokorina2017}.
Under the selected sea state conditions the wave variance and the ensemble averaged instant wavenumber spectrum do not change significantly in the course of the wave evolution; no wave breaking occurred.

\begin{figure}[b]
	\includegraphics[width=8.5cm]{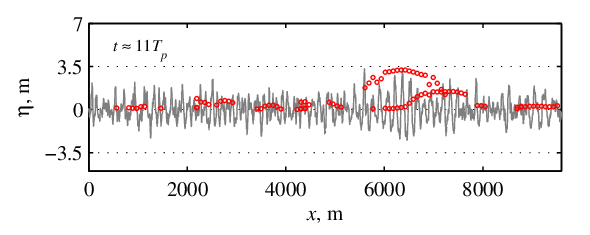}(a)
	\includegraphics[width=8.5cm]{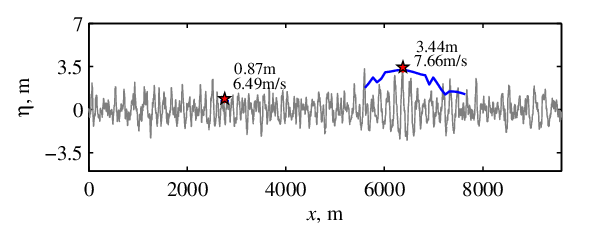}(b)
	\caption{\label{fig:Sample1} The wave surface in dimensional variables at $t \approx 11T_p$ and the results of two stages of the WIST, (a) and (b) respectively.}
\end{figure}

A snapshot of the surface at the moment not long after the beginning of the simulation, $t \approx 10 T_p$, is shown in Fig.~\ref{fig:Sample1} by the solid curve. Symbols in Fig.~\ref{fig:Sample1}a show the result of the first step of the WIST, when solitons in a short sliding window a sought, similar to Fig.~\ref{fig:EnvelopeSoliton}. One may see that many groups in the series are recognized as solitons, though there are only few of them which possess significant amplitudes. Two 'arcs' which reveal soliton-type groups may be seen in the interval between about $5500$~m and $8000$~m. At the second step, the 'arcs’ are analyzed by a dedicated subroutine, and locations of the biggest solitons are determined, see the blue curve with the red dot in Fig.~\ref{fig:Sample1}b. Then, the eigenvalue problem is solved once again in a larger interval around the selected solitons, what gives more accurate estimates (the stars in Fig.~\ref{fig:Sample1}b). The parameters of the revealed large-amplitude solitons (amplitudes $A_s$ and velocities $V_s$) are indicated in Fig.~\ref{fig:Sample1}b. 
At this instant, two large-amplitude soliton groups are recognized as the result of the WIST. The largest has the amplitude $A_{env} \approx A_s \approx 3.44$~m (which is close to the significant wave height $3.5$~m); its velocity is $V_s \approx 7.66$~m/s (which is close to the estimate of the linear wave group velocity $c_{gr} \approx 7.81$~m/s for 10-s waves). Though the second 'arc' in Fig.~\ref{fig:Sample1}a indicates the presence of another soliton to the right from the maximum one (at approximately $x = 7000$~m), it is not recognized by the automated procedure. According to Fig.~\ref{fig:Sample1}b, the second revealed soliton at $x \approx 2900$~m is of much smaller amplitude and propagates significantly slower. 
%This estimate is consistent with the appearance of the waves, which look significantly shorter than at the location of the maximum soliton. 
%
\begin{figure}[b]
	\includegraphics[width=8.5cm]{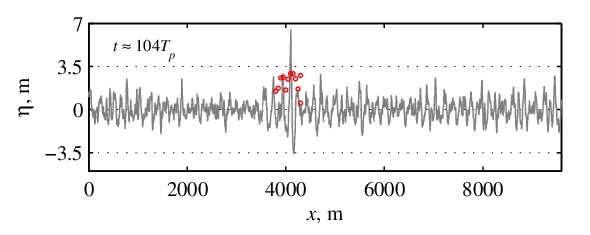}(a)
	\includegraphics[width=8.5cm]{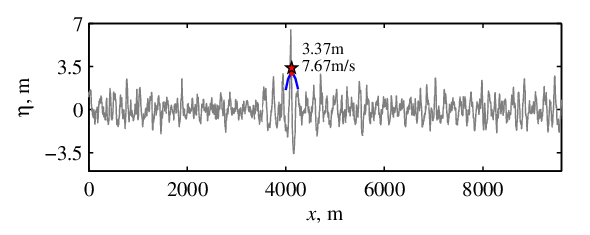}(b)
	\caption{\label{fig:Sample2} Same as in Fig.~\ref{fig:Sample1}, but for the instant $t = 1041$~s.}
\end{figure}
\begin{figure}[b]
	\includegraphics[width=8.5cm]{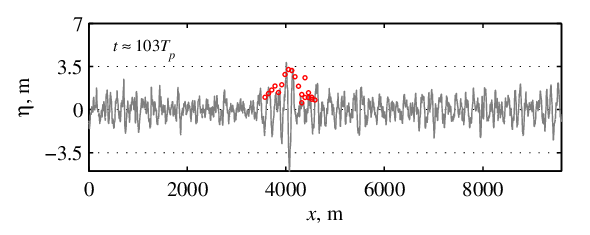}(a)
	\includegraphics[width=8.5cm]{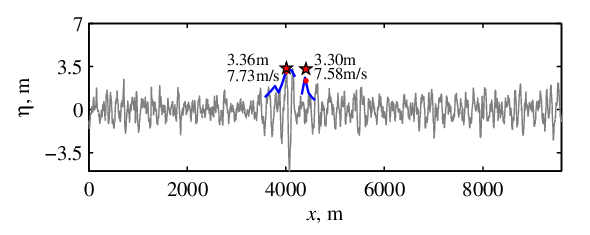}(b)
	\caption{\label{fig:Sample3} Same as in Fig.~\ref{fig:Sample1}, but for the instant $t = 1035$~s.}
\end{figure}

Two other examples how the WIST performs are shown in Fig.~\ref{fig:Sample2} and Fig.~\ref{fig:Sample3} for much later instants, after about 100 periods of the evolution. The case shown in Fig.~\ref{fig:Sample2} corresponds to the moment of one of the greatest elevations observed during the evolution. The case in Fig.~\ref{fig:Sample3} is about a half period earlier, when a very deep trough occurs. Note that in contrast to the case shown in Fig.~\ref{fig:Sample1}, single soliton groups are found at the first stage of the WIST. Hence, small-scale solitons seem to disappear in the course of the wave evolution. 
In general, the 'arcs' formed by the estimated soliton amplitudes look much shorter and rougher in Figs.~\ref{fig:Sample2},~\ref{fig:Sample3} compared to Fig.~\ref{fig:Sample1}. In accord, the wave group which is recognized as the soliton, looks much less regular than in Fig.~\ref{fig:Sample1}. 
At the same time, the estimated amplitudes and velocities of the maximum solitons in Figs.~\ref{fig:Sample1}, \ref{fig:Sample2}, \ref{fig:Sample3} are very close. In Fig.~\ref{fig:Sample3} the automated procedure has recognized two solitons with very similar parameters in close vicinity to each other. It is obviously the result of insufficient capability of the subroutine which determines locations of the solitons at the second stage, which erroneously interpreted the soliton group as two.

\begin{figure}[b]
	\includegraphics[width=8.5cm]{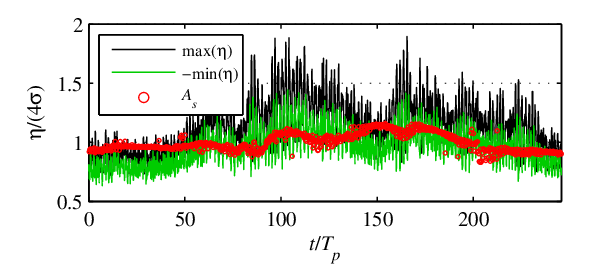}(a)
	\includegraphics[width=8.5cm]{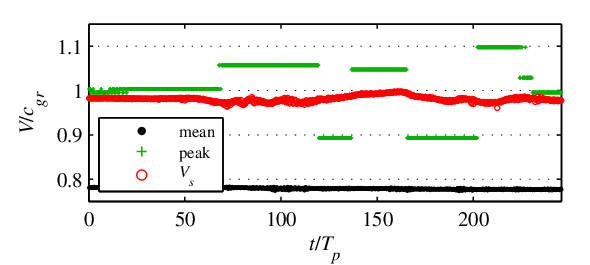}(b)
	\includegraphics[width=8.5cm]{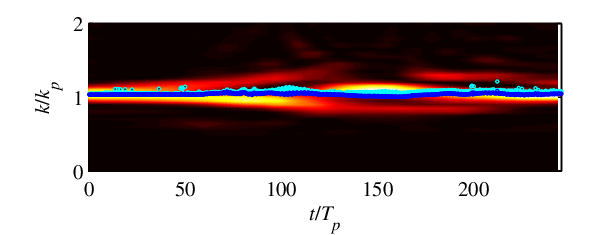}(c)
	\includegraphics[width=8.5cm]{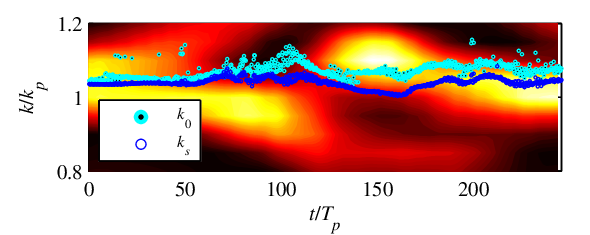}(d)
	\caption{\label{fig:SolitonAmplitudes} Evolution of the amplitude (a) and the velocity (b) of the maximum revealed soliton group. The maximum soliton amplitude $A_s$ is compared with the instantaneous wave extremes. The soliton velocity $V_s$ is compared with the peak and mean wave velocities calculated for the entire domain; $c_{gr}$ is calculated for the 10-s waves. The carrier wavenumbers  $k_0$ used by the IST procedure and the soliton wavenumbers $k_s$ are shown over the instantaneous spatial Fourier transform  of the short interval following the maximum wave group (c,d).}
\end{figure}

\begin{figure}[b]
	\includegraphics[width=8.5cm]{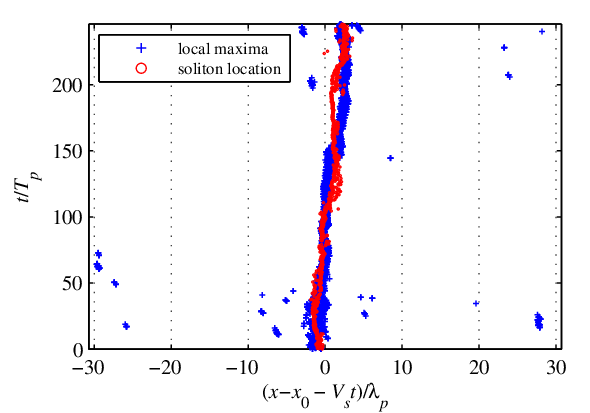}
	\caption{\label{fig:SolitonLocations} The same domain as in Fig.~\ref{fig:Surface1}, \textcolor{black}{but in the reference moving with the soliton velocity $V_s$. The } red circles show the location of the maximum soliton and the blue crosses mark the instant wave extremes, $\max{|\eta|}$.}
\end{figure}

The WIST is now applied to the frequent sequence of the surface snapshots, what allows to trace the evolution of the soliton groups. For simplicity, we consider only the soliton with the largest amplitude among all recognized solitons at a given time. The evolution of parameters of the largest soliton are shown in Fig.~\ref{fig:SolitonAmplitudes}. The amplitude  is shown with circles in Fig.~\ref{fig:SolitonAmplitudes}a. Note that (i)~the values of the soliton amplitude at the neighboring instants do not scatter much, what confirms robustness of the result of the WIST, and (ii)~the soliton amplitude is slowly changing, but does not alter much during about 240 wave periods (varies within about 17\%). These two observations are in accord with the data of the soliton velocity in Fig.~\ref{fig:SolitonAmplitudes}b; the velocity of the largest soliton varies within about 2\%. The locations of the maximum soliton at each instant are shown in Fig.~\ref{fig:SolitonLocations} with red circles. They correspond to the position of the intense wave group which is formed at the early stage of the evolution, thus this group remains the maximum soliton throughout the entire simulation period.

In Fig.~\ref{fig:SolitonAmplitudes}a the solid curves show the instantaneous maxima and minima of the surface displacement, $\max{\eta}$ and $-\min{\eta}$ respectively. It follows, that though within the first $50 T_p$ the values of the surface extremes are similar to the solitary group amplitude $A_s$ (about the significant wave height $4\sigma$), later on about twice larger waves occur (mainly due to the huge crests). The locations of the instant maximum displacements are shown in Fig.~\ref{fig:SolitonLocations} with blue crosses. One may see that the absolute majority of the extreme waves occur on top of the soliton group. Thus, the soliton group plays decisive role in the formation of extreme waves in this simulation.

In Fig.~\ref{fig:SolitonAmplitudes}b the linear wave group velocities calculated in the entire simulation domain basing on the wavenumber of the peak value of the momentary Fourier transform, and on the mean wavenumber, are shown by green crosses and black dots respectively. Due to the discreteness of the resolved wavenumbers, the peak wavenumber may jump between several competing Fourier modes. The mean wavenumber is stable during the evolution, it is significantly larger than the peak value, and hence the corresponding velocity is smaller than the others. 
As noted above, the soliton possesses almost constant velocity all the time; its wavelength roughly corresponds to the peak of the spectrum. Meanwhile, groups moving with different velocities may be readily seen in Fig.~\ref{fig:Surface1}. 

The evolution of the spatial Fourier transform of the surface within a short interval of $20\lambda_p$ following the maximum soliton is plotted in Fig.~\ref{fig:SolitonAmplitudes}c. The intensification and fading of a few different wave modes may be observed. The carrier wavenumbers $k_0$ used by the IST procedure are shown by cyan-black circles; they correspond to the local mean wavenumbers (better seen in the expanded scale in Fig.~\ref{fig:SolitonAmplitudes}d). The intrinsic wavenumber of the soliton $k_s$ may be calculated from the soliton velocity $V_s$ according to the dispersion relation (\ref{DispersionRelation}), $k_s = g/(2V_s)^2$. These values are plotted in Fig.~\ref{fig:SolitonAmplitudes}c,d with blues circles. 
One may see that the series of $k_s$ is much more stable than the local carrier wavenumbers $k_0$ and are slightly smaller in value. Note that the soliton wavenumber may correspond to none of the instant spectral peaks.
\textcolor{black}{In Fig.~\ref{fig:SolitonLocations} the frame of reference moves with the velocity $V_s$. It is obvious from the figure that the values $V_s$ underestimate the actual celerity of the soliton group.}

\section{\label{sec:Conclusions}Conclusions}

In this work we trace the evolution of a group of intense waves in the field of irregular deep-water waves with the help of the method based on the windowed Inverse Scattering Transform, WIST. 
The waves are characterized by the spectral conditions similar to the real sea, while the group retains its structure extraordinarily long. 
The method allows us to interpret waves in terms of envelope solitons of the nonlinear Schr\"odinger equation. %The present study of irregular waves generacontributes to the tests of the IST-based procedure performed in [Slunyaev, 2018]. 
The soliton parameters estimated in close time instants do not show significant scatter; they are determined with similar accuracy at the instants of appearance of huge crests (Fig.~\ref{fig:Sample2}) or deep troughs (Fig.~\ref{fig:Sample3}) despite the fact that the wave conditions are obviously beyond the formal limits of the NLS theory (waves are steep and the soliton group consists of just a few wave cycles). At the moment, the fully automated procedure may sometimes misinterpret the primary data of the windowed IST-based analysis, if several solitons interact or when the soliton group is extremely short. This issue requires further improvement.

%The simulated unidirectional irregular waves are essentially nonlinear and approximately satisfy the broad JONSWAP spectrum, hence poorly satisfy the assumptions of the NLS theory of small wave steepness and narrow spectrum. 

The wave evolution is calculated using the strongly nonlinear solver of the primitive potential water equations, hence is believed to be quite realistic. Surprisingly, an intense group is formed by chance from irregular waves and persists for remarkably long time. The WIST reveals the soliton nature of this group, and evaluates the key parameters of it, which are the intrinsic amplitude, velocity and location. The estimated amplitude and velocity of the soliton group vary a little during about 240 periods of the wave evolution, hence the hydrodynamic envelope soliton persists through nonlinear interactions with surrounding waves for so long. 
%At the same time, the effect of acquiring the wave energy by a large soliton known for nonintegrable systems (e.g. \cite{Mussotetal2009}) has not been observed either.
At the same time, the effect of an accruing soliton which borrows  the energy from smaller waves, known for nonintegrable systems (e.g. \cite{Mussotetal2009,Kachulinetal2020}), has not been observed.

Since the intrinsic amplitude of the soliton group is about the significant wave height and remains approximately constant in time, it is not much surprising that most of the extreme wave events occur on top of the soliton group and lead to heavier wings of the wave height PDF. The maximum surface displacement observed in the simulation is about two significant wave heights, $\max{\eta}/H_{1/3} \approx 1.9$, which is an extraordinary value. It may be interpreted as a superposition of the coherent soliton group with the remaining irregular wave background. This idea is in line with the discussion of in-situ rogue wave registrations in \cite{Slunyaev2006}, where a significant part of the recorded rogue waves was explained as the combination of soliton-like groups with the random background. In the present work we confirm possibility of this scenario of rogue wave generation using the direct numerical simulation.

The occurrence of long-living soliton-type patterns which facilitate the generation of extremely high waves should also lead to longer rogue wave events, in agreement with the direct evaluation of 3D rogue wave lifetimes reported in \cite{Fujimoto2019,KokorinaSlunyaev2019}.

Soliton-type groups and Peregrine breather-type wave patterns were found previously in water wave sequences by eye or by fitting in \cite{Hendersonetal1999,Clamondetal2006,Chabchoubetal2011,Cazaubieletal2018,AgafontsevGelash2020}, and also through the solutions of the associated scattering problem for the entire wave record, e.g. \cite{CaliniSchober2017,Osborneetal2019,Suretetal2020}.
%and by the analyzing the IST in a periodic domain [El]. 
There is also significant amount of relevant research in the field of nonlinear optics \cite{Akhmedievetal2016,Kibleretal2010,Tikan2020}. 
\textcolor{black}{The IST has recently been proposed as a tool for characterizing coherent structures which occur in systems which are not close to the ``parent" integrable model \cite{Chekhovskoyetal2019,Turitsynetal2020}.} 
However, to the best of our knowledge no one has observed truly soliton groups propagating through intense irregular broad-banded water waves for so long.
Hence in this work we underpin the interpretation of nonlinear modulated oceanic waves in terms of \textcolor{black}{solitons} by the solid basis.  

We should particularly mention the recent work \cite{Suretetal2020}, where ensembles of solitons were modeled in the experimental flume.  Though in these experiments the wave spectrum was narrow, the soltion gas was dense, and the distance of propagation was up to almost $100$ wave periods. Noticeable evolution of the scattering data (calculated for the entire wave series) was pointed out. Based on our previous analysis of the soliton content in in-situ rogue wave measurements \cite{Slunyaev2006}, and on the present work, we assume that the realistic sea conditions rather correspond to the situation when soliton-type groups are located sparsely.     

The planar geometry seems to be the major restriction of the present work. However, according to our preliminary study of three-dimensional waves, soliton-like wave patterns are able to survive for a least few tens of wave periods if the angle spectrum is not too broad (these results will be reported elsewhere). Thus, a wave group dynamics similar to the one presented in the this work can probably take place in the field of long-crested oceanic waves. 
%The occurrence of long-living soliton-type patterns which facilitate the generation of extremely high waves should also lead to longer rogue wave events, in agreement with the direct evaluation of 3D rogue wave lifetimes reported in \cite{Fujimoto2019,KokorinaSlunyaev2019}.

\begin{acknowledgments}
This research was funded by the RSF under Grant No. 16-17-00041; part of the research reported in Sec.~\ref{sec:IST} was supported by the  RFBR Grant No. 18-02-00042.
\end{acknowledgments}

\textbf{Data Availability Statement}

The data that support the findings of this study are available from the author upon reasonable request.

%\nocite{*}

\bibliography{HydroSolitonPersistence}% Produces the bibliography via BibTeX.

\end{document}